\documentclass[aps,prb,reprint]{revtex4-1}
\usepackage{amsmath}
\usepackage{amssymb}
\usepackage{units}
\usepackage{url}
\usepackage{amsfonts}
\usepackage{amssymb}
\usepackage{textcase}
\usepackage{bm}
\usepackage{natbib}
\usepackage{graphicx}
\usepackage{braket}
\usepackage{color}
\newcommand{\pvec}[1]{\vec{#1}\mkern2mu\vphantom{#1}}
\renewcommand{\vec}[1]{\boldsymbol{#1}}

\begin{document}
\title{Bielectrons in the Dirac sea in graphene: the role of many--body effects}
\author{L.L. Marnham}
\email{llm204@exeter.ac.uk}
\affiliation{School of Physics, University of Exeter, Stocker Road, Exeter EX4 4QL, United Kingdom}
\author{A.V. Shytov}
\affiliation{School of Physics, University of Exeter, Stocker Road, Exeter EX4 4QL, United Kingdom}

\begin{abstract}
It was shown in [\onlinecite{marnham15}] that the dynamics of a pair of electrons in graphene can be mapped onto that of a single particle with negative effective mass, leading to bound states of positive energy despite the formally repulsive interaction. However, this conclusion was based on the analysis of the two--particle problem, neglecting the role of the Dirac sea and the many--body effects. The two dominant such effects at zero temperature are screening of the Coulomb interaction by the Dirac sea, and reduction of the available phase space due to Pauli blocking of transitions into the states below the Fermi level. We show that these effects result in strong renormalization of the binding energy, but do not destroy the metastable states. Thus the binding energies are strongly dependent on the chemical potential owing to the combined effects of screening and Pauli blocking. Hence, the quasibound resonances can be tuned by electrostatic doping.
\end{abstract}
\maketitle
\section{Introduction}
\indent For over a decade, graphene has been the object of tremendous research attention due in part to the variety of novel pseudo--relativistic phenomena it made accessible in the condensed matter laboratory~\cite{katsnelson06,shytov07,gu11,beenakker06}.
One example of an open research problem in graphene is that of the zero--field excitonic condensate, which has seen significant treatment in the theoretical literature\cite{min08,zhang08,abergel2013}. The absence of the predicted experimental signatures~\cite{gorbachev12,kim11} has prompted some authors to study the prototypical two--body problem~\cite{marnham15,sabio10,lee12,mahmoodian13,downing15} in an attempt to better understand the emergent properties of these non--trivial correlated phases.\\
\indent Most of the interesting electronic behaviour in graphene can be attributed to its low energy bandstructure, which is effectively described by a pair of Dirac cones~\cite{wallace47} centered on the points $\vec{K}^+$ and $\vec{K}^-$ in the first Brillouin zone. To first order in momentum, the dispersion of a single electron near one of these points is given by $\epsilon=\pm v_F p$, where $v_F=10^6$ms$^{-1}$ is the Fermi velocity and $p=|\vec{p}|$ is the magnitude of the momentum relative to either $\vec{K}^+$ or $\vec{K}^-$. The dynamics of a pair of non--interacting electrons is described by the Hamiltonian:
\begin{align}
\widehat{H_L}=v_F\vec{\sigma}_1\cdot\hat{\boldsymbol{p}}_1+v_F\vec{\sigma}_2\cdot\hat{\boldsymbol{p}}_2,
\end{align}
where $\hat{\boldsymbol{p}}_i$ and $\boldsymbol{\sigma}_i$ are the momentum and pseudospin operators, respectively, and subscripts refer to particle $1$ and particle $2$. 
The most interesting two--particle behaviour occurs when the kinetic energy of the constituent particles is compensated, which occurs in the scattering channel $\vec{p}_1=-\vec{p}_2\equiv\vec{p}$. Indeed, $\widehat{H}_L$ has four energy configurations: $E_{1,2}=0$ (in which one electron is in the conduction band, and the other is in the valance band), $E_3=2v_Fp$ (when both electrons are in the upper cone) and $E_4=-2v_Fp$ (both electrons in the lower cone). The space of two--particle states is therefore partitioned into two important subspaces: the \textit{the non--dispersing sector}, which is spanned by states of zero energy, and the \textit{dispersing sector} of all other states. The two--particle eigenstates within the non--dispersing sector are given by\cite{marnham15}:
\begin{gather}\label{wf1}
\begin{aligned}
\ket{1,\phi_{\vec{p}}}&=\frac{1}{\sqrt{2}}\left[e^{-i\phi_{\vec{p}}}\ket{\uparrow\uparrow}+e^{i\phi_{\vec{p}}}\ket{\downarrow\downarrow}\right],\\
\ket{2,\phi_{\vec{p}}}&=\frac{1}{\sqrt{2}}\left[\ket{\uparrow\downarrow}+\ket{\downarrow\uparrow}\right],\\
\end{aligned}\raisetag{4\baselineskip}
\end{gather}
\noindent where the arrows represent the pseudospin configuration, in analogy with the problem of two spin--$\nicefrac{1}{2}$ particles, and $\phi_{\vec{p}}$ is the polar angle in momentum--space defined by $\tan(\phi_{\vec{p}})=\frac{p_y}{p_x}$. For the states spanned by $\ket{1,\phi_{\vec{p}}}$ and $\ket{2,\phi_{\vec{p}}}$, the kinetic energy compensation was shown in Ref. [\onlinecite{sabio10}] to lead to singular behaviour when interactions are introduced into the problem. However, the singularity arises only in the case of  \textit{exact} symmetry between the cones. A complete treatment of the problem therefore requires analysis beyond the conical approximation, e.g., by the introduction of a small band curvature.\\
\indent While the conical approximation to the dispersion is typically sufficient for describing the electronic properties of graphene, it was argued in Ref. [\onlinecite{marnham15}] that this is not true of the two--body problem. In the conical approximation, the kinetic energy of a single electron scales like $v_Fp\sim\frac{\hbar v_F}{r}$ due to the uncertainty principle. However, if the linear contributions cancel exactly (as they do for the energy configuration discussed above) the dominant term is due to a small band curvature, which instead scales like $p^2\sim \frac{\hbar^2}{r^2}$. If the particles interact via a Coulomb potential ($U(r)\sim\frac{e^2}{\epsilon r}$) then a quadratic kinetic energy of the appropriate sign\cite{marnham15} leads to bound state formation regardless of the interaction strength\cite{landau77}.\\
\indent In order to understand the origin and significance of the band curvature, we consider the single--particle Hamiltonian discussed in Ref. [\onlinecite{marnham15}], which is given by
\begin{gather}
\begin{aligned}
\widehat{H}_{j}&=v_F\vec{\sigma}_j\cdot\vec{p}_j-\frac{p^2_j}{4m^*}+\tau_j\mu (p_{x,j}+ip_{y,j})^2\sigma_{+,j}+\text{h.c.},\label{single}
\end{aligned}\raisetag{0.5\baselineskip}
\end{gather}
where h.c. denotes the Hermitian conjugate of the previous term, $\sigma_{+,j}=\frac{1}{2}\left(\sigma_{x,j}+ i\sigma_{y,j}\right)$ and $\tau_j=\pm$ determines the sign of the trigonal warping term for a particle near the $\vec{K}^{\pm}$--point. There are also two new parameters which determine the energy in Eq. (\ref{single}): the first is an effective mass $m^*=\frac{\hbar^2}{9a^2t'}$, which arises due to hopping between next--nearest--neighbours (where $t'$ is the next--nearest--neighbour hopping parameter \cite{castro09} and $a=1.42$\AA{} is the carbon--carbon distance in graphene); the second arises due to the trigonal warping of the bands, and is given by $\mu=\frac{3a^2t}{8\hbar^2}$ (where $t=2.8$eV is the nearest--neighbour--hopping parameter). The effective mass can be as high as $m^*=7.5m_e$ (see discussion of $t'$ below), so that the isotropic contribution to the band curvature is small. When $m^*<0.8m_e$ however, the curvature is sufficiently large to induce pairing\cite{marnham15}. Unlike the linear terms, the quadratic terms do not change their sign under $\vec{p}\rightarrow -\vec{p}$, and hence do not cancel in the non--dispersing sector ($E_{1,2}=0$). The effective Hamiltonian for particles in the non--dispersing sector can be constructed by projecting out the high energy states\cite{marnham15}:
\begin{align}
\widehat{H}_{1,2}^\text{eff}=\left[\begin{array}{cc}
-\frac{p^2}{2m^*} & \tau_{1,2}\mu p^2\sin(3\phi_{\vec{p}}) \\
\tau_{1,2}\mu p^2\sin(3\phi_{\vec{p}})  & -\frac{p^2}{2m^*}
\end{array}\right]\label{direct},
\end{align}
\noindent for two particles in the same valley, where the rows and columns correspond to states $\ket{1,\phi_{\vec{p}}}$ and $\ket{2,\phi_{\vec{p}}}$ and $\tau_{1,2}=\tau_1+\tau_2$. In this work we will consider the case of two particles in the same valley (the so--called \textit{direct pairs}) so that $\tau_{1,2}=\pm 2$.\\
\indent It was shown in Ref. [\onlinecite{marnham15}] that the electron--electron pair described by Eq. (\ref{direct}) has negative--definite kinetic energy when the nearest--neighbour--hopping parameter ($t'$) takes a certain size. We note that $t'$ is not tuneable, but there is some disagreement about its precise value (see the discussion in [\onlinecite{marnham15}] and references therein) which is expected to fall in the range $0.02t\leq t'\leq 0.2t$. The role of $t'$ can be understood by noting that Eq. (\ref{direct}) has eigenvalues of the form $E_{kin}=-2\mu p^2\left(\eta+\sin(3\phi_{\vec{p}})\right)$, where we have introduced the \textit{anisotropy parameter} $\eta=\frac{6t'}{t}$. If $\eta>1$ the isotropic term dominates, and the kinetic energy is negative definite, so that pairing will occur for a repulsive interaction. When $\eta<1$ the kinetic energy is sign--indefinite, and the anisotropy due to trigonal warping is more important\cite{marnham15}. The aforementioned range of possible values of $t'$ leads to a corresponding value for the anisotropy parameter: $0.12\leq \eta \leq 1.2$.\\
\indent It is instructive to compare the interaction effects described in Ref. [\onlinecite{marnham15}] to other, more familiar condensed matter systems. In semiconductors at low doping, attractive interactions result in binding between electrons near the bottom of an empty conduction band and holes near the top of a filled valence band, separated by a gap. When the concentrations of electrons and holes are sufficiently low, one can ignore the effects of Fermi statistics and describe the excitons in terms of two constituent massive particles. The resulting energy levels are dependent on dimensionality and the details of the interaction. In contrast to this, metals admit low--energy excitations which involve transitions between electron states near the Fermi level within the same band. Most of these transitions are blocked due to the Pauli exclusion principle, which may result in low--temperature anomalies and, eventually, the formation of strongly correlated phases such as ferromagnets, charge density waves, etc. The details of such transitions depend upon the interaction between the electrons, as well as the structure of the single particle spectrum. In the case of a weak attractive interaction and flat density of states, the restrictions arising from Pauli blocking can result in the formation of a bound state known as a Cooper pair, irrespective of the dimensionality. The many--body physics of graphene does not fit neatly into either of the scenarios above. On the one hand, monolayer graphene is gapless, so that elementary excitations cannot be thought of as entirely seperate from the sea of free carriers, as they can be in semiconductors. On the other hand, the density of states is linear, and vanishes near the charge--neutrality point. Therefore, graphene (near $E_f=0$) can not be considered to be a metal, either (for a full discussion see, e.g., Ref. [\onlinecite{kotov}]).  Further, due to the linear dispersion, the Coulomb coupling constant is independent of the density of free carriers, and electron--electron interactions are important irrespective of the doping. This is not the case for a two--dimensional electron gas with parabolic dispersion\cite{kotov}. It is therefore unclear whether the two--body problem will necessarily stand on its own or, e.g., be valid for only some (if any) doping regimes when the effects of the Dirac sea are taken into account. Thus, it is desirable to incorporate the effects of Pauli blocking and screening into the two--body problem of an interacting electron--electron pair.\\
\indent In this work, we show that the electron--electron pairs which form in the subspace of non--dispersing two--particle states have their energies renormalized by many--body effects. We study the two most important many--body phenomena at zero temperature: screening of the interaction by free carriers in the Dirac sea, and (ii) Pauli blocking, in which scattering into single--particle states below the Fermi level is forbidden due to the Pauli exclusion principle. We show that the problem admits an analytic solution when the interaction is short--ranged (in the subspace of states where the trigonal warping is compensated). We then treat the complete problem, with bandstructure warping and screened interaction potential, numerically. For both single-- and double--layered structures we find good agreement between numerical results and our exact solution which is applicable when the exponential factor is of the order of one, or even less.\\
\indent The rest of the paper will have the following structure. In Sec. II we discuss a simple heuristic model in which the subspace of states where trigonal warping is not important are studied. We show that this model has an analytic solution when the interaction potential is sufficiently short--ranged. In Sec. III we discuss the role of screening and examine the effective electron--electron interaction in single-- and double--layered structures. We show that the screening is vital for accurately determining the binding energies in systems near $E_F=0$, and exponentially suppresses the binding when the system is heavily doped. In Sec. IV we discuss the results of numerical calculations of the binding energies of pairs, including the effects of trigonal warping and momentum--dependent interaction neglected in Sec. II. We find that the numerical results agree with our analytical solutions in the low--doping regime, where the binding energies are proportional to the Fermi energy. We conclude in Sec. V with a summary of important results. In Appendix A we show how the two--particle Schr\"{o}dinger equation with the Pauli blocking constraint, Eq. (\ref{scheq}), can be derived from the Bethe--Salpeter equation, and in Appendix B we discuss the form of the electron--electron interaction for double layer systems with non--trivial dielectric environment.
\section{analytical solution for short--ranged interaction}
\indent A formal treatment of electron--electron pairing in the Dirac sea requires the solution of the Bethe--Salpeter equation. We perform that calculation in detail in Appendix A, but it is helpful to first treat a heuristic model which incorporates the most important many--body effects in a simple way.\\
\indent We begin with two natural assumptions: (i) that the free carriers screen the electron--electron interaction to the extent that it can be considered short--ranged, and (ii) transitions into paired states in which one of the electrons is below the Fermi level are blocked by the Pauli exclusion principle\cite{griffiths}. In the limit of large hole doping, the interaction is suppressed for distances $rp_f\gtrsim 1$, so as a first approximation we assume that the electron--electron interaction takes the form $V(\vec{r})=\lambda\delta\left(\vec{r}\right)$, where $\lambda$ determines the strength of the interaction which is different for single-- and double--layered systems. Our second assumption is equivalent to only allowing contributions of momenta in the range $0<p<p_F$. We will show that, subject to these assumptions, the problem admits an analytical solution.\\
\indent It is instructive to first recall the case of a two--dimensional massive particle in a short--ranged potential $V(\vec{r})$ of interaction radius $a$. For a weak potential, the bound state is shallow, and the wavefunction is concentrated at the distances $r\gg a$. Therefore, it is possible to approximate the potential by a delta function, $V(\vec{r})\approx \lambda\delta(\vec{r})$, where $\lambda=2\pi\int V(\vec{r})rdr$. This gives the binding energy in the form\cite{landau77}: 
\begin{align}
|E|\sim E_c\exp\left\lbrace-\frac{2\pi\hbar^2}{m\lambda}\right\rbrace. \label{approxenergy}
\end{align}
The prefactor $E_c\sim \hbar^2/ma^2$ represents the energy cutoff which is determined by the behaviour of $V(\vec{r})$ at distances $r\sim a$. Thus, it shows that the model with a delta--like potential is incomplete, and has to be regularized at short distances in a non--universal way. Therefore, Eq. (5) is only applicable for $\lambda\ll 2\pi\hbar^2/m$, so that the binding energy is determined by the universal exponentially small factor. Since the dynamics of the electron--electron pair is governed by quadratic kinetic energy, the bound state in this problem has similar properties. However the regularization is now provided by the momentum space cutoff at $p=p_F$, rather than the interaction radius $a$. We will show below that this leads to an exact solution for the binding energy of the bielectron, as opposed to the approximate solution of Eq. (\ref{approxenergy}). Not only does this cutoff extend the range of validity of the solution to $\lambda\gtrsim 2\pi\hbar^2/m$, we will show below that it is vital if one is to correctly determine the low--doping behaviour of the system.\\
\indent We consider the states with configuration $\ket{2,\phi_{\vec{p}}}=\frac{1}{\sqrt{2}}[\ket{\uparrow\downarrow}+\ket{\downarrow\uparrow}]$, so that the two--particle wavefunction is given by $\Psi=\sum\limits_{|\vec{p}|<p_F}\psi_p\ket{2,\phi_{\vec{p}}}$. In this case the kinetic energy is isotropic and there is a trivial overlap of states with different momenta\cite{marnham15}. Further, we assume that the deformation of the valence band due to trigonal warping is not important, so the Fermi surface can be approximated by a circle of radius $p_F$. Within the momentum representation, the Schr\"{o}dinger equation takes the form (from now on we choose $\hbar=1$):
\begin{align}
-\frac{p^2}{2m^*}\psi_{\vec{p}}+\iint\limits_{|\pvec{p}'|<p_F} U_{\vec{p}-\pvec{p}'}\psi_{\pvec{p}'}\frac{d^2p'}{(2\pi)^2}=E\psi_{\vec{p}}\label{scheq}.
\end{align}
The short--ranged potential is described by a momentum--independent interaction: $U_{\vec{p}-\pvec{p}'}=\lambda$. Within this approximation, the integral equation (\ref{scheq}) is separable, with eigenfunctions:
\begin{align}
\psi_p=\frac{\lambda}{2\pi}\frac{A}{E+\frac{p^2}{2m^*}}.\label{wavefunction}
\end{align}
Substitution of Eq. (\ref{wavefunction}) into Eq. (\ref{scheq}) gives a simple integral equation, which is readily solved:
\begin{align}
1=\frac{\lambda}{2\pi}\int_0^{p_F}\frac{p'dp'}{E+\frac{p'^2}{2m^*}}=\frac{\lambda m^*}{2\pi}\ln\left[\frac{p_F^2}{2m^*E}+1\right].
\end{align}
The binding energies of the pair therefore take the form:
\begin{align}
E=\frac{p_F^2}{2m^*}\times\frac{1}{\exp\left(\frac{2\pi}{m^*\lambda}\right)-1}.\label{energies}
\end{align}
For a repulsive potential, $\lambda>0$, the energies are positive definite. Further, the pairs are only destroyed by many--body effects in two limits: $p_F\rightarrow 0$ (the neutrality point, when the phase space is full and all pairing is blocked) and $\lambda\ll 2\pi/m^*$ (the limit of the ultra--dense hole gas, in which screening fully suppresses the interaction at all length scales). The states are therefore not destroyed for any systems of experimental interest.\\
\indent Eq. (\ref{energies}) can be understood as the product of two factors which compete to determine the binding energy. The first factor, which scales as $\sim p_F^2$, arose in our calculation due solely to the momentum cutoff in the integral equation (\ref{scheq}), and can therefore be understood as the contribution of Pauli blocking to the pairing. As the system is tuned towards the charge neutrality point from below, $E_F\rightarrow 0^-$, the phase space available to accommodate pairs in the channel $\vec{p}_1=-\vec{p}_2$ shrinks, and the binding is suppressed as $p_F\rightarrow 0$. The second factor depends on $\lambda$ and is therefore governed by the strength of the interaction. In the next section we will show that, $\lambda\sim p_F^{-n}$, so that for both cases of interest (monolayer ($n=1$) and double--layer ($n=2$) systems) it leads to an exponential suppression of the binding energy for large $p_F$.
\section{interaction screening}
\indent To understand the doping dependence of the binding energies from Eq. (\ref{energies}), we now discuss the dependence of the potential strength $\lambda\equiv\lambda(E_f)$ on the Fermi energy. We will derive the forms of the one-- and two--layer screened interaction within the random--phase approximation and use them to determine $\lambda$. We note that only the repulsive Coulomb interaction is considered in this paper. This is in contrast to, e.g., the case of traditional superconductivity, in which the repulsive Coulomb interaction is not sufficient to cause pair binding, and the Hamiltonian must be extended by the attractive electron--phonon interaction. Because the kinetic energy of the pair is negative--definite and parabolic in the momentum, binding occurs for an arbitrarily weak repulsive interaction\cite{marnham15} in two dimensions in analogy with the case of a massive particle in an arbitrarily shallow well, which has been shown to exhibit at least one bound state in, e.g., Ref. [\onlinecite{landau77}].\\
\indent We begin with the case of a graphene monolayer in an environment of dielectric constant $\kappa$. If $\Pi(q=0)\neq 0$, within the random phase approximation, the effective electron--electron interaction $U_{\vec{q}}$ takes the form
\begin{align}
U_{\vec{q}}=\frac{v_{q}}{1+\Pi(q) v_{q}},
\end{align}
where $\Pi(q)$ is the polarizability of the system and $v_q=2\pi e^2/\kappa q$ is the bare Coulomb interaction in the momentum representation. For sufficiently small momenta, $q\ll q_s$, the potential $U_q$ can be approximated by a short--ranged potential $V(\vec{r})=\lambda\delta(\vec{r})$, where $\lambda=U_{q=0}$ and $q_s$ is the inverse screening radius. The polarizability function (which was calculated within the random phase approximation in Ref. [\onlinecite{hwang07}]) is constant in the $q\leq 2p_F$ regime (we will suppress the momentum dependence and write $\Pi\equiv\Pi(q)$), and is used to obtain:
\begin{align}
\Pi v_q=\frac{N\alpha p_F}{q},
\end{align}
where $\alpha=e^2/\kappa v_F$ is the Coulomb coupling constant of graphene in a dielectric environment and $N$ is the number of fermion flavours \textit{per layer}. We can therefore write down the interaction strength for, $\lambda=U_{\vec{q}=0}$, for a single--layer system:
\begin{align}
\lambda=\frac{2\pi v_F\alpha}{q_s}\label{strength1},
\end{align}
where $q_s=N\alpha p_F$ is the Thomas--Fermi screening momentum of electrons in graphene.\\
\indent We see that as the doping increases the interaction strength $\lambda$ decreases, so that the binding becomes weak at large doping. Let us introduce the crossover scale by setting the argument of the exponential factor in Eq. (\ref{energies}) to unity. This occurs when the magnitude of the Fermi energy is of the order of $E_s\equiv m^*v_F^2/N\sim 1$eV. In the limit of high doping ($|E_F|\gg E_s$) the binding energies are exponentially suppressed:
\begin{align}
E\sim\exp\left(-\frac{|E_F|}{E_s}\right),
\end{align}
but we note that this suppression would be hard to observe, due to the large doping required: $|E_F|\gtrsim 1$eV. Hence, only the limit $|E_F|\ll E_s$ is discussed below.\\
\indent Let us now turn to the case of two layers separated by a dielectric spacer of thickness $d$. We assume that the device is fully embedded in the same dielectric material. The case of a dielectric spacer between the layers with permittivity different from that of the environment requires some extra analysis, which is given in Appendix B. Assuming the electrons are in opposite layers, the potential takes the form (see, e.g., Refs. [\onlinecite{profumo10}] and [\onlinecite{abergel2013}]):
\begin{align}
U_{\vec{q}}=\frac{v_q e^{-qd}}{\left(1+\Pi_1 v_q\right)\left(1+\Pi_2 v_q\right)-\Pi_1\Pi_2 v_q^2e^{-2qd}},\label{doublelayer}
\end{align}
where $\Pi_{1,2}$ are the polarizabilities of layer $1$ and $2$, respectively. In this work we restrict ourselves to the simplest case where the polarizabilities of the two layers are the same: $\Pi_1=\Pi_2\equiv\Pi$. This is equivalent to the assumption that the Fermi levels of the two layers are tuned to the same value: $E_{f,1}=E_{f,2}$. Following the reasoning applied to the single--layer case above, we obtain for the  interaction strength:
\begin{align}
\lambda=\frac{\pi v_F\alpha}{q_s(1+q_sd)}.\label{strength2}
\end{align}
Therefore, the crossover to weak binding in the double--layer case occurs at a different Fermi energy, $E_d$, which for inter--layer separation $d\gtrsim 1$nm is given by  $E_d=v_F/2\alpha Nd\sim 0.1$eV. The fact that $E_d<E_s$ is due to the large minimum separation of two particles when they are confined to separate layers: even when the in--plane distance between the electrons is zero, they are still separated by $d\sim 1$nm. In the limit of high doping ($|E_F|\gg E_d$) the binding energies are again exponentially suppressed:
\begin{align}
E\sim\exp\left(-\frac{E_F^2}{E_sE_d}\right),
\end{align}
We therefore expect screening to be the limiting factor in the binding energy for large $p_F$.\\
\indent On the other hand, the Coulomb interaction is only weakly screened in graphene doped near the neutrality point, $|E_F|\ll E_d\ll E_s$. As noted above, Pauli blocking is particularly important in this regime, giving a contribution $\sim p_F^2$. Nevertheless, screening does reduce this suppression for finite negative doping, leading to binding energies which are linear in the Fermi energy:
\begin{align}
E=\frac{p_F^2}{2m^*}\cdot\frac{m^*\lambda}{2\pi}=\frac{1}{2Nn}|E_F|,\label{linear}
\end{align}
where $n=1,2$ is the number of graphene layers. We note the role of the exact cutoff momenta of Eq. (\ref{scheq}): they prevent the binding energies from becoming singular as $E_F\rightarrow 0^-$.\\
\indent Finally, let us reiterate that we have focussed our attention on the simplest case of equal doping, when $\Pi_1=\Pi_2$. Because the pairs reside in opposite cones and have opposite momenta, the limiting role is played by the layer which hosts the lower--cone electron (which we will now call layer 1, without loss of generality). Clearly Fermi statistics does not change the problem at all for any level of hole doping in layer 2, as the necessary phase space is always free for an electron in the upper cone: all that is required is $E_{f,2}\leq 0$. On the other hand, the form of Eq. (\ref{doublelayer}) suggests that inequivalent doping might significantly change the strength of the interlayer potential. Let the Fermi energies of layers $1$ and $2$ be given by $E_F$ and $E_F+\delta E_F$, respectively. The interaction strength is then given by:
\begin{align}
\lambda(\delta E_F)=\frac{\lambda(0)}{1+\frac{2\delta E_F}{|E_F|}}.
\end{align}
We therefore conclude that our results are relevant for a range of doping regimes, so long as $|E_{f,1}-E_{f,2}|\ll |E_{f,1}|$.
\section{The effects of finite--range interaction and trigonal warping}
\begin{figure}[h]
\includegraphics[width=\linewidth]{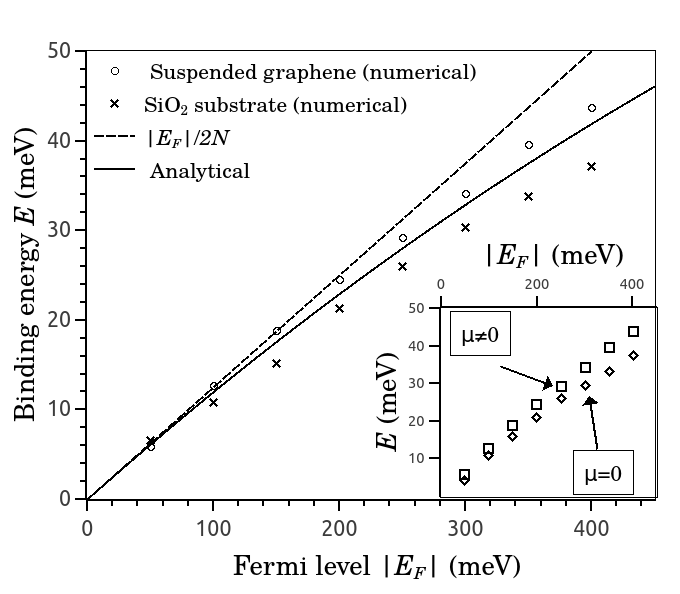}
\caption{Binding energies in single--layer systems. The symbols represent numerical results, the solid line is the analytical form of Eq. (\ref{energies}) and the broken line corresponds to the low--doping approximation of Eq. (\ref{linear}). The inset shows numerical results for a single--layer suspended graphene structure with trigonal warping switched on ($\mu\neq 0$) and off ($\mu=0$). The small increase in kinetic energy from the trigonal warping explains the difference between numerical results and the analytical approximation (which was calculated with isotropic kinetic energy) for the case of suspended graphene.}
\label{fig_results1}
\end{figure}

\begin{figure}[h]
\includegraphics[width=\linewidth]{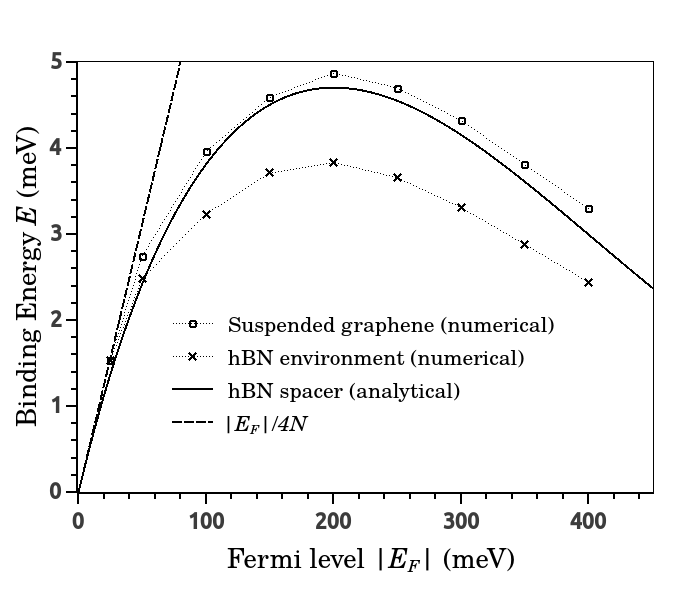}
\caption{Binding energies in double--layer (graphene--hBN--graphene) structures which are suspended over vacuum (circles) or embedded in an hBN environment (crosses). The solid line is the analytical result for the double--layered system embedded in hBN, and the broken line is the low--doping approximation.}
\label{fig_results2}
\end{figure}

\indent The analytic form of the binding energies, along with the low--doping approximation, are plotted in Figs. \ref{fig_results1} and \ref{fig_results2} for the cases of single-- and double--layer devices, respectively. For the two--layer case, the results were obtained for graphene in a hexagonal boron nitride (hBN) environment. For the one--layer case, however, our analytical result is entirely independent of the substrate (see Eq. \ref{strength1}). In both configurations the linear approximation is substrate--independent, confirming that the low--doping behaviour is dominated by momentum--space restrictions due to Pauli blocking. The points on those plots correspond to binding energies obtained numerically (see below), and account for contributions from the full non--dispersing sector, as well as the momentum--dependent interaction. We have used $\eta=1.1$ throughout.\\
\indent So far we have considered a subspace of the possible bound states of the system: those with configuration $\ket{2,\phi_{\vec{p}}}=\frac{1}{\sqrt{2}}[\ket{\uparrow\downarrow}+\ket{\downarrow\uparrow}]$. It was shown that the isotropic dispersion ($\propto -p^2/2m^*$) leads to an exact solution under the assumption that the potential is short--ranged. When we consider other states in the non--dispersing sector, however, the effects of trigonal warping enter the problem in two ways: (i) there is an extra contribution to the kinetic energy (see Eq. (\ref{direct})), and (ii) the Fermi surface is deformed, so that states excluded by Pauli blocking occupy a domain with a more complicated shape (see Appendix A for further details). In addition, we treated an electron--electron potential of the form $V(\vec{r})=\lambda\delta(\vec{r})$, which tends to overestimate the interaction strength. In Ref. [\onlinecite{marnham15}] it was shown that the effective Hamiltonian of an electron pair in the non--dispersing sector is given by
\begin{align}
\widehat{H}_{\vec{p},\pvec{p}'}=\delta_{\vec{p},\pvec{p}'}\widehat{H}_{1,2}^\text{eff}
+U_{\vec{p},\pvec{p}'}&\left[\begin{array}{cc}
\cos(\phi_{\vec{p}}-\phi_{\pvec{p}'}) & 0 \\
0 & 1\label{hamiltonian}
\end{array}\right],
\end{align}
due to the non--trivial overlap of wavefunctions with different momenta: $\braket{1,\phi_{\pvec{p}'}|1,\phi_{\vec{p}}}=\cos(\phi_{\vec{p}}-\phi_{\pvec{p}'})$. When the effects of screening (so that $U_{\vec{p},\pvec{p}'}\equiv U(|\vec{p}-\pvec{p}'|)$ is given by the potentials in Sec. III) and the warped Fermi surface are taken into account, the problem becomes intractable. Therefore, in this section we present the results of the numerical diagonalization of Eq. (\ref{hamiltonian}), and compare them with the analytical solution found above.\\

\indent For monolayer systems, we assume the graphene lies on a substrate of dielectric constant $\epsilon_s$, so that $\kappa=\frac{1}{2}\left(1+\epsilon_s\right)$. We present results for two common experimental systems: (i) suspended graphene $\left(\epsilon_s=1\right)$ and (ii) graphene on a SiO$_2$ substrate $\left(\epsilon_s=3.9\right)$. For double--layer systems the situation is slightly different, as the graphene layers must be seperated by a dielectric spacer in order to suppress bilayer coupling. We provide results for a hexagonal boron nitride (hbN) spacer. hBN has dielectric constant $\epsilon_h=3.9$ and has been experimentally shown to electrically isolate parallel graphene layers at a thickness of 4 atomic layers ($d=1.3$nm)\cite{britnell12}. We have considered two distinct double--layered cases: hBN--graphene--hBN structures which are (i) embedded in a hBN environment ($\kappa=\epsilon_h$) and (ii) suspended in vaccuum ($\kappa=\frac{1}{2}\left(1+\epsilon_h\right)$).\\
\indent For the double--layer case (see Fig. \ref{fig_results2}) there is good agreement between the numerical results and the linear approximation of Eq. (\ref{linear}) for $|E_F|\lesssim 50$meV. For higher doping, however, the analytical solution overestimates the binding energy. Our assumption that the Fermi surface is circular (i.e. ignoring trigonal warping) is asymptotically exact in the limit $p_F\rightarrow 0$ so we expect our analytical solution from Sec. I to diverge from the numerical results (which take account of warping) for large $p_F$. We also note that the interaction potential $U_q$ is positive definite and has a maximum at $q=0$. Therefore, using the interaction strength $\lambda=U_{q=0}$ is guaranteed to overestimate the strength of the potential. The curve is peaked at $E_F\sim -200$meV. This is due to the small value of $E_d\sim 0.2$eV, which defines an energy scale accessible even at relatively low doping.\\
\indent For the single--layer case, the linear approximation of our analytical result agrees well with numerical values over the whole region of experimental interest ($|E_F|\lesssim 400$meV). In contrast to the double--layer case, the exponential suppression of the binding is not accessible in this range, however, due to the relatively large energy scale of the single--layer problem: $E_s\sim 5E_d$.\\
\indent We also note that the single--layer numerical results of Fig. (\ref{fig_results1}) for the case of suspended graphene are slightly higher than the analytic result. This is entirely due to trigonal warping, which was not accounted for in Eq. (\ref{scheq}). The effective Hamiltonian of Eq. (\ref{direct}) has eigenvalues of the form $\epsilon=-2\mu p^2\left[\eta+\sin(3\phi_{\vec{p}})\right]$, which correspond to the kinetic energy of the pair. Although this term is negative definite for $\eta>1$, the kinetic energy is suppressed for certain momentum space configurations due to the sign--indefinite trigonal contribution. This leads to increased binding energy when interactions are taken into account, an effect not accounted for in the purely isotropic Hamiltonian used in the analytic approximation of Sec. II. The inset of Fig. \ref{fig_results1} displays the binding energies of a bielectron in single--layer suspended graphene with ($\mu\neq 0$) and without ($\mu=0$) trigonal warping switched on. The results show that trigonal warping tends to increase the binding energy of the pair, as discussed above. We note that the results for $\mu=0$ and $\mu\neq 0$ diverge from each other as the Fermi energy is decreased. This is to be expected, as warping of the energy bands is only appreciable at high energies. When the system is doped near the neutrality point, the difference between phase--space restriction for the warped and conical cases is negligible.\\
\indent Although we have obtained numerical results up to $E_F=400$meV, we note that such high doping may not be practical for experiment. For the cases of monolayer graphene on SiO$_2$ and double layer graphene in a hBN environment, the limiting factor is likely to be dielectric breakdown in the region between graphene and gate. For the example of hBN, it was found in tunnelling experiments\cite{britnell12} that breakdown occurs at $3$V for a 4--layer--thick spacer, so that it is possible to achieve $|E_F|\sim 800$meV. Due to the large breakdown field of vacuum, other factors are expected to limit the range of validity for suspended graphene. When not supported by a substrate, finite charge densities of opposite sign on the gate and graphene respectively lead to deflection of the graphene away from its usual atomically flat structure\cite{bolotin08,bunch07,fogler08}, which can result in the electrostatic collapse of the device. As an example, we consider the device of Ref. [\onlinecite{bolotin08}]: a square graphene sample of dimension $\sim 3\mu$m is suspended $150$nm above a $150$nm thick layer of SiO$_2$ (which rests atop a Si substrate). When a gate voltage $V_G$ was applied between graphene and Si, collapse occurred at $V_G=20$V. This corresponds to a maximum doping of $|E_F|\sim 90$meV. The precise upper bound which is experimentally accessible depends strongly on device geometry, which we do not consider here.\\ 
\section{conclusions}
\indent In this paper, we have considered the problem of an electron--electron pair in single-- and double--layered graphene structures, taking into account the most important effects at zero temperature: interaction screening and the blocking of single--particle transitions to states below the Fermi level by the Pauli exclusion principle. We have not, however, considered the effects of finite temperature which may be important, particularly for the case of double--layered structures where the binding energies are rather low ($0$meV--$5$meV). The positivity of the binding energies implies that the corresponding non--trivial correlated phase would be rather unusual: unlike the case of the more familiar excitons, which have negative binding energies due to an attractive interaction, the pairs described in this paper have positive energies. Hence, their formation is not energetically favourable. It was shown in Ref. [\onlinecite{marnham15}], however, that their decay rates are slow. Therefore, the bielectronic condensate, should it form, would be metastable, and would not be a candidate for novel superconductivity.\\
\indent We have shown that the binding energies are renormalized, but not destroyed, by the presence of the Dirac sea. Within the range of attainable doping, the binding energies are about an order of magnitude higher for single--layered structures, suggesting that they are of the greatest experimental interest. In particular we have shown that, for sufficiently low hole doping ($|E_F|\lesssim 400$meV for one layer and $|E_F|\lesssim 50$meV for two layers) the renormalized energies are directly proportional to the magnitude of the Fermi energy, to a good approximation. We therefore conclude that the pairing is tunable by a gate voltage for single and double layer systems, and that the results are robust against a small doping mismatch between the layers.\\
\section{acknowledgements}
\indent The authors wish to thank V.I. Fal'ko, M.V. Berry and C. Downs for insightful discussions. A.V.S. is supported by EPSRC/HEFCE No. EP/G036101.
\appendix
\section{Derivation of the effective two--body Schr\"{o}dinger equation}
To incorporate the effects of the Dirac sea into the problem, we wish to solve the Bethe--Salpeter equation:

\begin{equation}
\includegraphics[width=0.87\linewidth]{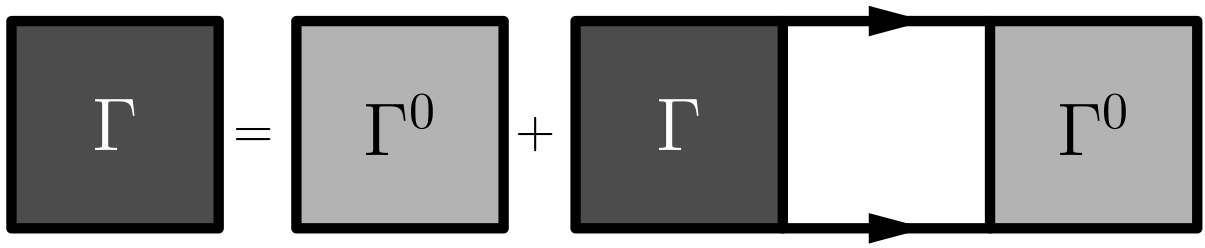}
\label{bsequation}
\end{equation}

\noindent where the solid lines represent free electron propagators, $\Gamma$ is the electron--electron scattering amplitude representing a sum over ladder diagrams, and $\Gamma^0$ is the sum over all irreducible diagrams:
\begin{equation}
\includegraphics[width=0.87\linewidth]{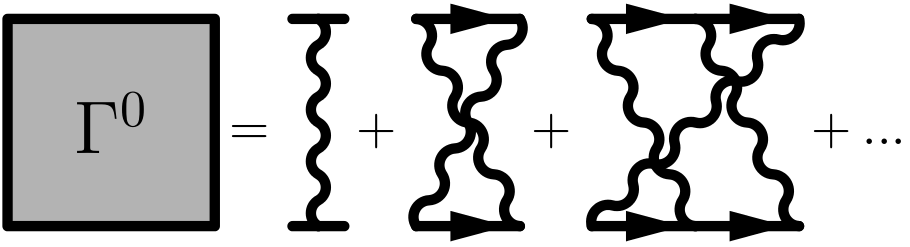}
\label{irreducible diagrams}
\end{equation}
To leading order in momentum, electrons in graphene travel with velocity $v_F=c/300$, and so it is sufficient to ignore retardation effects and treat the interaction as instantaneous (i.e., the dielectric function takes the form $\epsilon(\vec{q},\omega)=\epsilon(\vec{q},0)$). In this case, all diagrams in Eq. (\ref{irreducible diagrams}) containing crossed interaction lines involve frequency integrals with poles in the same half plane for particles of the same species, and therefore vanish exactly. Therefore, there is only one irreducible part with non--trivial contribution in Eq. (\ref{irreducible diagrams}), which takes the form $\Gamma^0\sim U_{\vec{p}-\pvec{p}'}$.\\
\indent In the momentum representation, Eq. (\ref{bsequation}) gives the following integral equation, which is to be solved for $\Gamma$:
\begin{align}
&\Gamma(\pvec{p}',\vec{p};\vec{K}\omega)=U_{\vec{p}-\pvec{p}'}+\nonumber\\
&\int U_{\pvec{p}''-\pvec{p}'}\left[i\int G_1(\pvec{p}'',\epsilon)G_2(\vec{K}-\pvec{p}'',\omega-\epsilon)\frac{d\epsilon}{2\pi}\right]\times\nonumber\\
&\Gamma(\pvec{p}'',\vec{p};\vec{K}\omega)\frac{d\pvec{p}''}{(2\pi)^2}.\label{gamma2}
\end{align}
The momenta in Eq. (\ref{gamma2}) are interpreted as follows: $\vec{p} (\pvec{p}')$ is the incoming (outgoing) momentum of particle 1, and $\vec{K}$ is the total momentum. We note that $\vec{K}$ is a constant of the motion, due to momentum conserving factors in the interaction matrix element $U_q$, which arise due to translation invariance of the real--space potential: $V(\vec{r}_1,\vec{r}_2)=V(|\vec{r}_1-\vec{r}_2|)$. Thus, $\vec{K}-\vec{p}\text{  }(\vec{K}-\pvec{p}')$ is the incoming (outgoing) momentum of particle 2. In the body of this work, we have focussed on the $\vec{K}=0$ channel, but we will delay this restriction until the end for generality.\\
\indent Because we assume the Fermi level is away from charge neutrality, we must modify the Green's function so as to account for some states with $E<0$ being vacant. It should also be noted that, for holes, the Green's function is that of an electron moving backwards in time which amounts to the replacement $\delta\rightarrow -\delta$. We will therefore switch to the language of the so--called causal Green's functions, which account for states above ($\omega>E_F$) and below ($\omega<E_F$) the Fermi level. The Green's function of particle $i=1,2$ in the chiral representation becomes:
\begin{align}
G_i(\vec{p},\omega)&=\sum_{\lambda}\frac{\widehat{P}_{i,\lambda}}{\omega+E_F-E_{i,\lambda}+i\delta_{\vec{p},\lambda}^{(i)}},\label{anomalous}
\end{align}
where, for $E_F<0$ and at zero temperature:
\begin{align}
\delta_{\vec{p},1}^{(i)}&=\delta_{\vec{p},2}^{(i)}=\begin{cases}
\delta, & p<p_F, \\
-\delta, & p>p_F,\\
\end{cases}\nonumber\\
\delta_{\vec{p},3}^{(i)}&=\delta_{\vec{p},4}^{(i)}=\delta,\label{infinitesimal1}
\end{align}
with $\delta=0^+$. The index $\lambda$ labels the eigenstates of the single particle Hamiltonian from Eq. (\ref{single}), and $E_{i,\lambda}$ is the corresponding eigenvalue. $\widehat{P}_{i,\lambda}$ is the projection operator onto the state $\ket{i,\lambda}\otimes\ket{\psi_{j\neq i}}$ in the space of particle $i$.\\
\indent After substitution of the Green's function given by Eq. (\ref{anomalous}), the frequency integral in Eq. (\ref{gamma2}) takes the form:
\begin{align}
\frac{i}{2\pi}\int_{-\infty}^{\infty} G_1(\epsilon)G_2(\omega-\epsilon)d\epsilon=\frac{i}{2\pi}\sum_{\lambda,\lambda'}\widehat{P}_{1,\lambda}\widehat{P}_{2,\lambda'}\int_{-\infty}^{\infty}\frac{d\epsilon}{h(\epsilon)},\label{contour}
\end{align}
where $h(\epsilon)=(\epsilon+E_F-E_{1,\lambda}+i\delta_{\lambda}^{(1)})(\omega-\epsilon+E_F-E_{2,\lambda'}+i\delta_{\lambda'}^{(2)})$. The integrand has two poles, both of which can be on either side of the real axis depending on the value of the energy $E_{i,\lambda}$. We choose for a contour the semicircle which follows the  $\operatorname{Re}(\epsilon)$--axis from $-\infty$ to $\infty$ and completes with an arc of infinite radius by noting that the integral vanishes over the arc. As we can choose to complete the contour in the upper (lower) half--plane, all terms in Eq. (\ref{contour}) with both poles below (above) the real axis vanish. Therefore, the contour integral is carried out term--by--term, giving:
\begin{align}
&\Gamma(\pvec{p}',\vec{p};\vec{K}\omega)=U_{\vec{p}-\pvec{p}'}+\nonumber\\
&\int\frac{U_{\pvec{p}''-\pvec{p}'}\widehat{N}_{\pvec{p}'',\vec{K}}}{\omega+2E_F-\widehat{H}_{1}(\pvec{p}'')-\widehat{H}_2(\vec{K}-\pvec{p}'')+i\widehat{N}_{\pvec{p}'',\vec{K}}\delta}\times\nonumber\\
&\Gamma(\pvec{p}'',\vec{p};\vec{K}\omega)\frac{d\pvec{p}''}{(2\pi)^2},\label{gamma3}
\end{align}
where we have defined $\widehat{N}_{\pvec{p}'',\vec{K}}=\widehat{1}(1-n_{1}(\pvec{p}'')-n_{2}(\vec{K}-\pvec{p}''))$ from the Fermi--Dirac distributions at zero temperature:
\begin{align}
n_i(p)=\begin{cases}
0 & p<p_F\text{  ;  }E_i<0, \\
1 & p>p_F\text{  ;  }E_i<0,\\
0 & \forall p\text{  ;  }E_i>0.\\
\end{cases}
\end{align}
In order to map the integral equation (\ref{gamma3}) onto the two--particle Schr\"{o}dinger equation, we propose the following change of variables
\begin{align}
\Gamma(\pvec{p}'',\pvec{p};\vec{K}\omega)=\int U_{\pvec{Q}'}\widehat{N}_{\pvec{p}''-\pvec{Q}',\vec{K}}\Psi(\pvec{p}''-\pvec{Q}',\vec{p};\vec{K}\omega)\frac{d\vec{Q}'}{(2\pi)^2}\label{psi2}.
\end{align}
which we take to be a definition of $\Psi$ rather than a solution for $\Gamma$. Substitution into Eq. (\ref{gamma3}) gives an integral equation in $\Psi$ which we solve with the ansatz:
\begin{align}
&\Psi(\pvec{p}',\vec{p};\vec{K}\omega)=(2\pi)^2\delta^{(2)}(\vec{p}-\pvec{p}')+\nonumber\\
&\frac{1}{\omega+2E_F-\widehat{H}_1(\pvec{p}')-\widehat{H}_2(\vec{K}-\pvec{p}')+i\widehat{N}_{\pvec{p}',\vec{K}}\delta}\times\nonumber\\
&\times\int U_{\pvec{q}'}\widehat{N}_{\pvec{p}'-\pvec{q}',\vec{K}}\Psi(\pvec{p}'-\pvec{q}',\vec{p};\vec{K}\omega)\frac{d\pvec{q}'}{(2\pi)^2}.\label{ansatz}
\end{align}
\indent Let us examine the significance of Eq. (\ref{ansatz}). The frequency $\omega$ gives the energy of the pair, and $E_f$ simply sets the zero point of the energy, so we can make the replacement $\omega+2E_f\equiv E$. The first term in Eq. (\ref{ansatz}) is energy independent, while the second term has poles which correspond to two--particle states. Near these poles, the first term can be dropped. Further, the integrand vanishes due to $\widehat{N}$ whenever the particle in the valance band is below the Fermi level. We can therefore absorb it into the limits of the integral:
\begin{align}
&\left(\widehat{H}_1(\pvec{p}')+\widehat{H}_2(\vec{K}-\pvec{p}')\right)\Psi_{\pvec{p}',\vec{K}}+\nonumber\\
&\iint\limits_{|\pvec{q}'+\pvec{p}'|<p_F} U_{\pvec{q}'}\Psi_{\pvec{p}'-\pvec{q}',\vec{K}}\frac{d\pvec{q}'}{(2\pi)^2}=E\Psi_{\pvec{p}',\vec{K}}.\label{solution}
\end{align}
In the $\vec{K}=0$ channel, Eq. (\ref{solution}) is equivalent to the two--particle Schr\"{o}dinger equation we presented as a heuristic model in Sec. II when one projects the Hamiltonian onto the states of configuration $\ket{2,\phi_{\vec{p}}}$, such that $\Psi=\sum\limits_{|\vec{p}|<p_F}\psi_p\ket{2,\phi_{\vec{p}}}$. This justifies the assumption of Sec. II in the absence of trigonal warping. To include the warping effects, all four states have to be included in the model. The states $\ket{3,\phi_{\vec{p}}}$ and $\ket{4,\phi_{\vec{p}}}$, however, are less important than $\ket{1,\phi_{\vec{p}}}$ due to energy mismatch.\\
\section{Interlayer interactions in a non--trivial dielectric environment}
\begin{figure}[h]
\includegraphics[width=\linewidth]{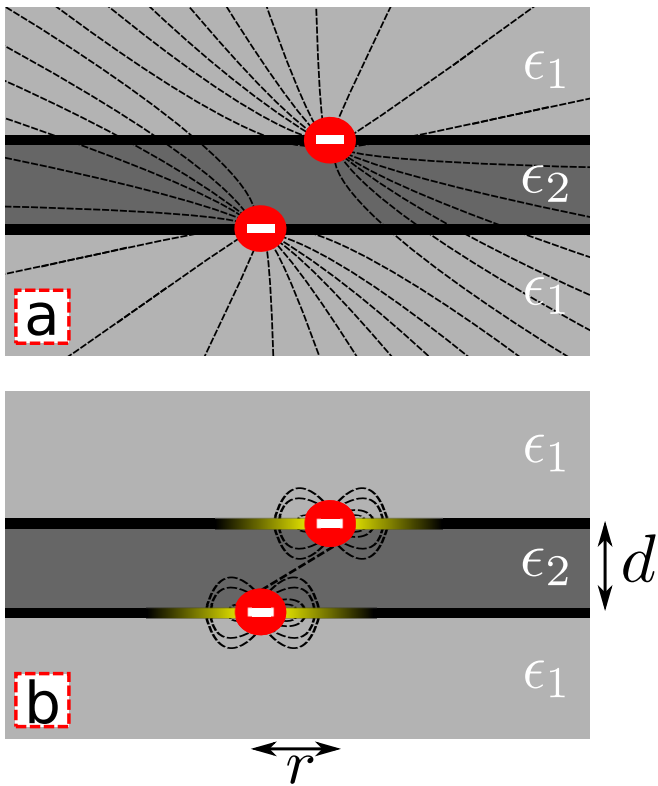}
\caption{(Color online) Schematic of the inequivalent dielectric double--layer. Two graphene layers are seperated by a dielectric spacer of permittivity $\epsilon_2$ and thickness $d$. The double layer is encapsulated in a second dielectric material, which has permittivity $\epsilon_1$. The layers host a pair of electrons (red circles) which have in--plane separation $r$. a) In the absence of screening by the electron sea, most of the electric field lines pass through the outer dielectrics and the interaction is mostly determined by the dielectric constant of the environment, $\epsilon_1$. b) When screening is considered, the electron charge is compensated by a positive screening cloud (yellow) within each graphene layer. The majority of the field lines now terminate on the screening charge, and the resulting (weak) interaction propagates through the middle dielectric. Hence, the interaction becomes sensitive to the value of the dielectric constant of the spacer, $\epsilon_2$.}
\label{fig_device}
\end{figure}

Let us now consider the case where the inter--layer spacer (with dielectric constant $\epsilon_2$) is a different material to the encapsulating dielectrics (which have dielectric constant $\epsilon_1$). The device in question is shown in Fig. (\ref{fig_device}), for $\epsilon_1\neq\epsilon_2$. In Sec. III, we discussed the form of the screened electron--electron interaction for double--layer structures. In that analysis, it was assumed that the dielectric material used to encapsulate the two layers was the same as that which separated them (i.e. $\epsilon_1=\epsilon_2$ in Fig. (\ref{fig_device})). In this appendix we calculate the interaction strength which is applicable to the results (see Sec. IV) for suspended double--layered systems, which corresponds to $\epsilon_1=1$ (the role of the outer dielectrics being played by vacuum) and $\epsilon_2=3.9$ (for a hBN spacer). We find that the interaction strength is governed by the permittivity of the spacer, and has the same form as it did in the $\epsilon_1=\epsilon_2$ case (see Eq. (\ref{strength2})).\\
\indent The screened inter--layer interaction, which has been considered previously by several authors (for more details see, e.g., Refs. [\onlinecite{profumo10}],[\onlinecite{badalyan13}] and [\onlinecite{hosono14}]), takes the form:
\begin{align}
U_q=\frac{v_{12}}{(1+v_{11}\Pi_1)(1+v_{22}\Pi_2)-v_{12}v_{21}\Pi_1\Pi_2},\label{screenedpotential}
\end{align}
where $v_{ii}$ is the bare intra--layer interaction for two particles in layer $i$ and $v_{ij}$ is the bare inter--layer interaction between electrons in layers $i$ and $j$. The bare interactions $v$ are given by:
\begin{align}
v_{11}&=v_{22}=\frac{2\pi e^2}{q}\times\frac{f_1}{f_2},\nonumber\\
v_{12}&=v_{21}=\frac{2\pi e^2}{q}\times\frac{\epsilon_2}{f_2},\label{barevs}
\end{align}
where:
\begin{align}
f_1&=\exp(qd)\left[\frac{\epsilon_1+\epsilon_2}{2}\right]-\exp(-qd)\left[\frac{\epsilon_1-\epsilon_2}{2}\right]\nonumber,\\
f_2&=\exp(qd)\left[\frac{\epsilon_1+\epsilon_2}{2}\right]^2-\exp(-qd)\left[\frac{\epsilon_1-\epsilon_2}{2}\right]^2.\label{f1f2}
\end{align}
We note that the substitution of the definitions in Eq. (\ref{barevs}) and (\ref{f1f2}) into the screened potential, Eq. (\ref{screenedpotential}), gives the form of the potential used in Sec. III when $\epsilon_1=\epsilon_2$ (see Eq. (\ref{doublelayer})).\\
\indent It is not immediately obvious which of the dielectrics is most important in determining the screening behaviour. Taking the limit of $q\rightarrow 0$, we obtain the interaction strength:
\begin{align}
\lambda=U_{q=0}=\frac{\pi e^2}{q_b+\frac{q_b^2 d}{\epsilon_2}},
\end{align}
where $q_b=N\alpha_b p_f$ is a kind of ``bare screening radius" determined by the bare coupling constant: $\alpha_b=e^2/v_F$. Thus, the material used to encapsulate the double--layered structure does not play a role in determining the strength of the momentum--independent interaction $V(r)=\lambda\delta(r)$. We therefore define the Coulomb coupling constant to be $\alpha=e^2/\epsilon_2v_F$, which in turn implies that $q_s=q_b/\epsilon_2$, and obtain
\begin{align}
\lambda=\frac{\pi v_F\alpha}{q_s(1+q_sd)},\label{strengthinequiv}
\end{align}
for $q_sd\gg 1$. This brings $\lambda$ to the form given by Eq. (\ref{strength2}), so that $\epsilon_2$ does indeed play the role of the effective dielectric constant.\\
\indent Naively, one might expect the interaction strength to be determined by the outer dielectrics, $\epsilon_1$, because it is through these layers that most of the electric field lines pass in the case of bare interactions (see Fig. \ref{fig_device}a). In fact, the result arises due to metallic screening of the interaction in the two graphene layers, which leads to charge compensation within a screening cloud of radius $1/q_s\sim1$\AA{}, so that the in--plane separation of the two electrons is very small: $r\ll d$. Because the charge is compensated, almost all electric field lines will begin on one electron and terminate on its surrounding ionic cloud. Occasionally, however, an electric field line will ``leak out" leading to a relatively weak coupling mediated through the inter--layer region, of dielectric constant $\epsilon_2$ (see Fig. \ref{fig_device}b). The two electrons therefore interact directly through the middle dielectric. This can be seen by taking the limit $d\rightarrow 0$ in Eq. (\ref{strengthinequiv}), which formally removes the spacer from the problem. In that limit, $\lambda$ does not depend on either of the two permittivities.

\bibliographystyle{unsrt}
\end{document}